\pdfoutput=1
\documentclass[preprint,aps]{revtex4}
\begin{document}
\title{Control of the Onset of Filamentation in Condensed Media}
\author{A. K. Dharmadhikari, K. Alti, J. A. Dharmadhikari, and D. Mathur}
\affiliation{Tata Institute of Fundamental Research, 1 Homi Bhabha Road, Mumbai 400 005, India.}
\date{\today}
\begin{abstract}
Propagation of intense, ultrashort laser pulses through condensed media like crystals of BaF$_2$ and sapphire results in the formation of filaments. We demonstrate that the onset of filamentation may be controlled by rotating the plane of polarization of incident light. We directly visualize filamentation in BaF$_2$ via six-photon absorption-induced fluorescence and, concomitantly, by probing the spectral and spatial properties of white light that is generated.   
\end{abstract}
\maketitle

\section{Introduction}

Filamentation is a spectacular nonlinear optical process that is induced by ultrashort pulses of intense light in matter. Its importance stems from the fact that optical energy becomes precisely localized within a filament.  Filamentation is important in the macroscopic domain: in air it has been used to control natural processes like lightening and is of utility in remote sensing and in light detection and ranging (see \cite{physrep} for a recent, cogent review). The ability to localize optical energy in bulk media also opens tantalizing possibilities in diverse applied areas, like micromachining. An important ramification of the localization of optical energy within ocndensed media lies in the posibility of making localized modification of material properties like refractive index, thus opening up applications like 3-D material modification and fabrication {\it within} bulk materials without surface damage. The feasibility of ``optically fabricating" photonic devices like resonators, beam-shapers, interferometers and amplifiers has already been demonstrated \cite{Schaffer}. Periodic arrays of nanostructures have been reported in fused silica \cite{rb}. Light-induced microexplosions within crystals have been shown to generate shock and refraction waves which, in turn, create nanovoids that have applications in emerging areas like nanophotonics and plasmonics \cite{Juodkazis}. 

From a fundamental viewpoint, the propagation of ultrashort light pulses in bulk media is complex, and interesting, as both the temporal and spatial profiles of the incident pulse are altered by the combined action of linear and nonlinear effects like group velocity dispersion (GVD), linear diffraction, self-phase modulation (SPM), self-focusing, multiphoton ionization (MPI), plasma defocusing and self steepening \cite{Alfano}. White light generation is another facet of propagation effects wherein additional frequency components are added to the incident spectrum. Recent work \cite{ourwork} indicates that SPM is responsible for spectral broadening that is symmetric around the incident wavelength; at higher powers, mechanisms like self-steepening and MPI-induced free-electron production generate blue-side frequency components, resulting in asymmetric broadening. 

The physics governing filament formation in transparent media is yet to be fully understood. Prevailing wisdom has, in the main, been based on dynamics that are driven by the interplay between self-focusing, diffraction, and defocusing due to free electron generation \cite{Lange,Brodeur}. New insights have recently begun to be obtained using the X-waves approach \cite{xwaves} wherein space-time propagation dynamics after self-focusing are characterized by pulse splitting that leads to the formation of waves that possess an {\bf  X}-shape both in the near field (when measurements are made in the space-time domain) and in the far field (wavenumber-frequency domain); this approach has succeeded in connecting theory with observations.     

Can some measure of control be exercised over filamentation dynamics? An answer in the affirmative would be of obvious importance, and constitutes the subject of this work. We report here on the control of the onset of filamentation by rotating the plane of polarization of incident light on crystals like BaF$_2$ (15 mm length) and sapphire (3 mm). We monitor filamentation by imaging the fluorescence signal obtained upon six-photon absorption of incident laser light in BaF$_2$. Such visualization yields direct evidence that control over the angle of polarization of the incident light enables us to readily exercise spatial control of filamentation within the crystal. Control is also verified by concomitantly monitoring the spatial and spectral properties of white light that is generated. We have also carried out experiments on 3 mm long crystals like sapphire. We emphasize here that the work that we report was carried out purely using linearly polarized laser pulses. We note that complementary work has been carried out earlier on polarization-dependent propagation dynamics and white light generation, but all such work has relied on altering the polarization state from linear to circular \cite{physrep,kolesik} and observing the changes in subsequent dynamics. Advantages of our approach vis-\'a-vis control over the onset of filamentation are discussed.

\section{Experimental method}

We used a Ti-sapphire laser yielding 36 fs pulses (820 nm wavelength) at 1 kHz repetition rate at incident powers of 6-300 $\mu$J, corresponding to $\sim$10-1000 times the critical power for self-focusing. A 30 cm lens loosely focused the incident light on to the target srystals and a $\lambda$/2 plate was used to vary the angle of polarization of the laser light. The resulting white light was imaged on a screen that was placed 50 cm from the crystal; a CCD camera captured the image. We characterized the spectra using an integrating sphere attached to a fibreoptic-coupled spectrometer (wavelength range 400-1100 nm) \cite{ourwork}. Filamentation was imaged using a digital camera. Both barium fluoride and sapphire crystals were oriented in the [100] direction, as confirmed by Laue back-reflection. We rotated each crystal around the light propagation axis so as to maximize the generation of white light. This defined the axis around which our $\lambda$/2 plate was rotated.   

BaF$_2$ allows direct visualization of the self-focusing/defocusing events by six-photon absorption-induced emission. The band gap of BaF$_2$ is 9.1 eV and our incident photon energy was ~1.5 eV. The 6-photon-induced blue-colored filaments imaged in Fig. 1 vividly demonstrate that energy localization within the crystal is readily amenable to spatial control: we spatially translated the filaments by rotating the polarization vector, keeping the incident power constant and we quantified polarization-controlled motion in terms of z$_{of}$, the distance from the entrance face of the crystal beyond which we observe the 6PA-induced emission; this is a robust indicator of the filament-start position. z$_{of}$ has contributions from physical focusing by the external lens and from self-focusing due to n$_2$, the intensity-dependent nonlinear refractive index of BaF$_2$. We fixed the position of the external lens and rotated the $\lambda$/2 plate. 

\section{Results and discussion}
\subsection{Filamentation in BaF$_2$}

Some typical images of filamentation formation within BaF$_2$ are depicted in Fig. 1. In this data set the distance between the lens and crystal was 28 cm and the geometrical focus of the lens was located outside the crystal. With the $\lambda$/2 plate at 0$^o$, z$_{of}$=3.5 mm (vertical white line in Fig. 1). As the incident polarization angle rotates there is gradual spatial translation of the filament-start position. At 45$^o$ the filament-start position lies well within the crystal, with z$_{of}\sim$5.9 mm. The value of z$_{of}$ reverts to its original 3.5 mm when the $\lambda$/2 plate is at 90$^o$. Figure 2 shows the functional dependence of how z$_{of}$ varies with polarization angle. 

Can the physical shift of z$_{of}$ be rationalized within the context of existing knowledge? To probe this, we take recourse to methods used in related numerical studies of light-induced damage in bulk media wherein z$_{of}$ has been related to parameters \cite{Junnarkar} like focal length, beam size, the distance between the lens and the entrance face of the crystal, the incident laser power and P$_{cr}$, the critical power for self focusing which, for isotropic materials, is (3.77$\lambda^2$)/(8 $\pi$n$_o$n$_2$). We note that P$_{cr}$ depends inversely on n$_2$ which contributes to the net refractive index of a material: n = n$_o$ + n$_2|E|^2$, where n$_0$ is the linear refractive index. n$_2$ is further related to the material's third-order susceptibility tensor: n$_2$ = (12$\pi$/n$_o$)$\chi^{(3)}$. In the ultrashort regime that our measurements have been conducted in, only electronic nonlinearities contribute to n$_2$. A pure electronic mechanism involves no translation of nuclei or rotation of atomic centers, and has relaxation times much less than the optical period (1/$\omega_o$). In case of cubic crystals like BaF$_2$ with space-group symmetry $m3m$, the effective value of $\chi^{(3)}$ is taken by us to be \cite{Payne}:
\begin{equation}
\chi^{(3)}(\theta) = {3 \chi^{(3)}_{1122} + (\chi^{(3)}_{1111} - 3 \chi^{(3)}_{1122})}{{[{\rm cos^2}(2\theta) + 1]}\over{2}},
\end{equation}
where $\theta$ is the incident polarization angle. As $\theta$ changes, the effective value of $\chi^{(3)}$ also changes and this, in turn, alters n$_2$. Thus, P$_{cr}$ is direction dependent. We made an estimate of z$_{of}$ following the treatment of \cite{Junnarkar} using these experimental parameters: incident power 100 P$_{cr}$, distance between sample and lens = 28 cm, and beam radius = 170 $\mu$m. Assuming that the incident power remains constant, by varying the value of P$_{cr}$ we can estimate the change in z$_{of}$. The theoretically expected functional dependence of z$_{of}$ in BaF$_2$ on parameters like P$_{cr}$ and incident laser power are depicted in Fig. 3. Using eqn.(1), it becomes clear that in order to account for the experimentally observed change in z$_{of}$, the required change in P$_{cr}$ would be as much as six times the initial P$_{cr}$. We estimate that the change in n$_2$ for BaF$_2$ is only 0.43$\times$10$^{-16}$ cm$^2$ W$^{-1}$ as $\theta$ changes by 45$^o$, resulting in only a 130\% change in P$_{cr}$. We have, of course, assumed that the effective value of $\chi^{(3)}$ remains a useful and valid parameter in the ultrashort regime where the polarization state may change on timescales that are comparable to our pulse duration. If that be the case, one may need Stokes parameters at a number of time intervals within a single laser pulse. However, as is clear from our experimental data, factors other than n$_2$ may be at play that need to be identified and explored. 

We note that nonlinear absorption effects will not affect the beam power at the onset of filamentation because these will set in only after sufficient intensity has been attained. Indeed, the beam power may actually decrease with propagation due to GVD-induced pulse broadening. However, this effect does not significantly broaden the pulse as z$_{of}$ changes from 3.5 mm to 5.9 mm. Other factors that we have considered include refractive index modification, color center generation, and induced birefringence. It is known that crystals like BaF$_2$ possess intrinsic birefringence only in the ultraviolet region (at wavelengths less than 365 nm) \cite{burnett}. Consequently, intrinsic birefringence can safely be ignored in the present experiments. The loose focusing employed in our experiments rules out the possibility of modifying refractive index, generating color centers or melting \cite{melting}. However, we note that loose focusing might well give rise to cross-polarized wave generation \cite{Jullien} wherein a single input wave is partly converted to a wave that is polarized perpendicular to the input plane polarization by degenerate four wave mixing. To probe this we measured the intensity of transmitted white light as a function of BaF$_2$ crystal rotation over the range of incident energies 6-20 $\mu$J. Results shown in Fig. 4 pertain to measurements we made at incident energy of 20 $\mu$J and clearly depict self-induced polarization change: the transmitted intensity depends on the induced phase shift and angle between the crystal axis and the incident polarization vector and can be theoretically predicted \cite{midorikawa,stolen}. The transmission T is theoretically given by \cite{stolen}
\begin{equation}
T = sin^2(\Delta\phi_{nl}/2)sin^2(2\theta),
\end{equation}
where $\Delta\phi_{nl}$ = (2$\pi$L/3$\lambda$)n$_2$I(cos$^2\theta$ - sin$^2\theta$). Here $\Delta\phi_{nl}$ is the induced phase shift between the electric field components along two orthogonal crystal axes, n$_2$ is the nonlinear index coefficient, L is the length of medium, I is the incident intensity, and $\lambda$ is the wavelength of the incident laser light. In the calculated data shown in Fig. 4, the incident intensity was taken to be 0.2 TW cm$^{-2}$ (based on our measurement of incident energy and the pulse duration, assuming Gaussian beam focusing conditions), L = 15 mm, and the value of n$_2$ was taken to be 1.7$\times$10$^{-16}$ cm$^2$W$^{-1}$ \cite{richard}. Our experimental results clearly show that minima appear every 45$^o$, in accord with the calculated transmission.

To place our polarization-angle-dependent control results in perspective, we note that in other recent work it has been possible to achieve a measure of control on filament morphology in air and glass by forcing aberrations in the incident light beam and by means of diaphragms or phase masks \cite{Dubietis}. M\'echain {\it et al.} \cite{Dubietis} have shown that by varying the laser power the self-focusing distance in air can be altered. We also made measurements at various incident powers. At relatively low power a single filament is obtained whose start position lies well inside the crystal. As the power increases, the filament-start position moves nearer the entrance face of the crystal. We found that in order to effect a change in filament-start position of $\sim$2.4 mm, as much as a 4-fold increase in incident laser power was required. 

\subsection{White light generation in BaF$_2$ and sapphire}

Recent studies have explored the efficiency of white light production and of its spectral content as a function of incident laser energy \cite{ourwork} but, to our knowledge, studies of how the overall shape of the spectra change as a function of polarization angle have not been reported. We now discuss the spectral dependence on the rotation of linearly-polarized light. White light generation depends on the length of filamentation and on the value of effective $\chi^{(3)}$ which, in turn, also has a polarization angle dependence.  Figure 5a shows polarization-angle-dependent white light spectra in BaF$_2$. The 820 nm light was blocked and only the white light was recorded. Symmetric and asymmetric components are clearly visible. Data depicted in Fig. 5b show that there is significant polarization-dependent spectral change, specifically a reduction in the spectral width as the incident laser polarization nears 45$^o$. 

For uniaxial crystals like sapphire, the relation for effective $\chi^{(3)}$ becomes more complicated as there is now a direction-dependence for $e$-rays but not for $o$-rays \cite{Chi}. The value of P$_{cr}$ will vary accordingly. In sapphire, where filamentation was not visualized due to the absence of 6PA-emission, we used white light spectra as the diagnostic. White light spectra from sapphire (see Fig. 6a) have two components \cite{ourwork}: one due to SPM that arises from the Kerr nonlinearity (symmetric about the incident wavelength) and an asymmetric component that arises from MPI-induced free electron generation. The latter contribution is on the blue side of the spectrum and in all our measurements we saw a polarization-dependent change in this component. The incident 36 fs pulse underwent splitting, with the $e-$, $o-$pulses propagating with different group velocities and the splitting ratio depended on the incident polarization. The group velocity mismatch length is readily estimated to be $\sim$3 mm for sapphire using the approach given in standard texts \cite{yariv}. Data depicted in Fig. 6a show that there is major polarization-dependent spectral change, specifically a reduction in the spectral width as the incident laser polarization nears 45$^o$. We note that the blue part of the spectrum is the result of contributions made by MPI-generated free electrons to the nonlinear, time-varying refractive index \cite{ourwork} and the self-steepening process \cite{physrep}. The functional dependence of the area under the white light spectrum on polarization angle is depicted in Fig. 6b. We note the similarity in this functional dependence with that depicted for BaF$_2$ in Fig. 5b.

In case of crystals possessing intrinsic birefringence (sapphire) and induced birefringence (BaF$_2$), with the  $\lambda$/2 plate at 0$^o$ or 90$^o$, the white light spectra are more or less identical. At exactly 45$^o$, the amplitudes of the $e$ and $o$-pulses are equal. Taking GVM-induced pulse splitting into account each pulse undergoes self-focusing at different times but with half the incident power. The spectra that we measured at 0$^o$, 20$^o$, and 40$^o$ are very different, a reflection of GVM-induced splitting. The reduction in blue component of the white light spectrum sensitively maps the lessening of effective laser intensity within the medium which is driven by alterations in self-focusing conditions that, in turn, arise from GVM-induced pulse splitting. 

Over and above measurement of spectra, it is also instructive to directly visualize the white light spot. Results for BaF$_2$ irradiated at different polarization angles (Fig. 7) demonstrate another manifestation of filament translation. When the filament-start is close to the crystal entrance the white light spot is large. As the half-wave plate is rotated, the filament-start position moves inside the crystal, reducing the white spot in the far field. We observed more than a two-fold reduction in the white light spot size when the polarization angle was rotated by 45$^o$. Figure 7 also shows evidence of concentric rings; this conical emission is a manifestation of spatio-temporal modulation instability and can be rationalized using the X-waves approach \cite{xwaves2}. As the polarization angle alters the ring contrast degrades and the conical emission is reduced: in the radial direction only those frequencies that satisfy the phase matching condition form concentric rings. Similar results were observed for sapphire.

We have demonstrated a simple method of controlling the position of the filament-start position by altering the plane of polarization of the intense light that is incident on crystals possessing intrinsic birefringence (sapphire) and in which birefringence is induced (BaF$_2$) \cite{George}. The major contribution to shifts in the filament-start position is found to be from GVM-induced pulse splitting. For BaF$_2$, when the power reduces by half the filament-start position shifts by 1.2 mm (compared to the measured shift of 2.4 mm). 

\section{Summary}

In our experiments we have directly monitored propagation dynamics in condensed media by imaging filaments in barium fluoride by virtue of 6-photon absorption of incident light by the crystal. Such direct visualization of filaments offers a ready handle with which to assess different methods of exercising control over the filament dynamics. Hitherto, such control could only be exercised by varying the input power, with obvious limitations to the extent to which control can be exercised without causing damage to the optical medium through which propagation is being studied. Our contribution is to vary the plane of polarization of the incident radiation to exercise control, and our filament imaging provides a ready means to assess the measure of control that we achieve. It turns out that a very much wider `dynamic range' of control can be exercised using the polarization vector. 

But what about optical media in which filament imaging is not possible, such as sapphire? We relate filament properties to the other facet of ultrashort propagation dynamics in condensed media, namely white light generation. We qualitatively map filament dynamics to the spatial and spectral properties of the white light that is generated. Consequently, we are able to assess the efficacy of control of filament onset by also monitoring the properties of the white light continuum. 

Our findings have diverse ramifications in applied areas as well as in the basic sciences, particularly from the viewpoint of material modification in the bulk. 

\acknowledgements

Discussions with A. Couairon and See-Leang Chin are acknowledged. We also thank A. Thamizhavel for crystal characterization. The Homi Bhabha Fellowship Council is thanked for financially supporting one of us (JAD). The Department of Science and Technology is thanked for partial financial support for the ultrafast laser system used in these experiments.

\newpage

\begin{figure}
\caption{Dependence of the spatial location at which filamentation commences in a BaF$_2$ crystal on the plane of polarization of incident light. The horizontal white line marked 15 mm indicates the spatial extent of the crystal. Vertical white lines are a guide to the shift in filament-start position with polarization angle.}
\end{figure}

\begin{figure}
\caption{Functional dependence of z$_{of}$ measured in BaF$_2$ on the incident polarization angle.}
\end{figure}

\begin{figure}
\caption{Theoretically expected functional dependence of z$_{of}$ in BaF$_2$ on parameters like critical power for self-focusing, P$_{cr}$, and incident laser power (see text for parameters used in the calculations).}
\end{figure} 

\begin{figure}
\caption{Variation of white light intensity with BaF$_2$ crystal rotation angle (dots) shows self-induced birefringence. The incident laser energy was 20 $\mu$J.  No external polarizer was used and the incident laser's extinction ratio (ER) was measured to be 700:1. A Glan-Thompson polarizer with ER=10$^5$:1 was used as an analyzer. Solid line is a theoretical prediction (see text).} 
\end{figure}

\begin{figure}
\caption{a) Polarization-dependent white light spectra obtained upon irradiation of a 15 mm long BaF$_2$ crystal (incident energy = 140 $\mu$J). b) Polarization dependent full width at half maximum (FWHM) of white light spectra from BaF$_2$.}
\end{figure}

\begin{figure}
\caption{a) Polarization-dependent white light spectra obtained upon irradiation of a 3 mm long sapphire crystal (incident energy = 50 $\mu$J). b)Polarization dependent area under the white light spectrum measured in sapphire.} 
\end{figure}

\begin{figure}
\caption{Image of the white light spot in BaF$_2$ as a function of polarization angle (16 $\mu$J). A screen was placed 50 cm from the crystal and the screen image was captured by a CCD camera. The diameter of the central, white light spot was measured by counting pixels.}
\end{figure}


\begin{references}
\bibitem{George}Other experiments on isotropic materials have shown that at high intensities induced birefringence makes any isotropic material behave like a uniaxial crystal.
\end{references}
\end{document}